\documentclass[aps,prl,twocolumn,superscriptaddress,showpacs,amsmath,floatfix,citeautoscript]{revtex4-2}
\usepackage{amssymb}
\usepackage{amsmath}
\usepackage{colordvi}
\usepackage{amsthm}
\usepackage{subeqnarray}
\usepackage{cases}
\usepackage{enumerate}
\usepackage{graphicx}
\usepackage{epsfig}
\usepackage{epstopdf}
\usepackage{subfigure}
\usepackage{color}
\usepackage{ulem}
\usepackage{xcolor}
\usepackage[english]{babel}

\begin{document}

	\title{Topological Origin of Intrinsic High Chern Numbers in Two-Dimensional M$_2$X$_2$ Materials}
	\author{Zujian Dai}
	\affiliation{Laboratory of Quantum Information, University of Science and Technology of China, Hefei 230026, China}
    \affiliation{Institute of Artificial Intelligence, Hefei Comprehensive National Science Center, Hefei, 230088, China}
	\author{Xudong Zhu}
    \affiliation{Institute of Artificial Intelligence, Hefei Comprehensive National Science Center, Hefei, 230088, China}
	\author{Lixin He}
	\email{helx@ustc.edu.cn}
	\affiliation{Laboratory of Quantum Information, University of Science and Technology of China, Hefei 230026, China}
	\affiliation{Institute of Artificial Intelligence, Hefei Comprehensive National Science Center, Hefei, 230088, China}
	\affiliation{Hefei National Laboratory, University of Science and Technology of China, Hefei 230088, China}

\begin{abstract}
Despite sharing a common lattice structure, monolayer M$_2$X$_2$ compounds realize quantum anomalous Hall phases with distinct Chern numbers, a striking phenomenon that has not been fully exploared.
Combining first-principles calculations with symmetry analysis and tight-binding models, we identify two generic band-inversion mechanisms governed by the orbital composition and symmetry representations of 3$d$ states near the Fermi level. When $d_{xz}/d_{yz}$ orbtials dominate, a doubly degenerate $\Gamma$-point inversion yields $C=1$; otherwise, inversions occur along $\Gamma$–X and $\Gamma$–Y at four $C_4$-related momenta, whose Berry-curvature contributions add to give $C=2$, distinct from scenarios relying on multiple bands inversions at a single $\mathbf{k}$ point. The same mechanism consistently explains related two-dimensional systems, including LiFeSe, KTiSb, MgFeP, and Janus M$_2$X$_2$ derivatives. The mechanism provide practical guidance for screening and engineering tunable high-Chern-number insulators.
\end{abstract}

	\maketitle
	
The quantum anomalous Hall (QAH) state~\cite{Haldane_1988,fangzhong_2010,Wenghongming2015,qixiaoliang_2016,xueqikun_2018,tokura_2019,MacDonald_2023} is a two-dimensional insulating phase characterized by an integer-quantized Hall resistance and vanishing longitudinal resistance, even in the absence of an external magnetic field. It is described by a nontrivial bulk topological Chern number $C$~\cite{TKNN_1982}, and is therefore also referred to as a Chern insulator.
Chern insulators have been realized in diverse two-dimensional platforms, including magnetically doped (Bi,Sb)$_2$Te$_3$ films~\cite{xueqikun_2013,changcuizu_2015,kouxufeng_2014,xueqikun_2018_exp}, intrinsic MnBi$_2$Te$_4$~\cite{zhangyuanbo_2020,wangyayu_2020,xuyong_2020,wangyayu_2021}, and moiré graphene and TMD heterostructures~\cite{graphene_qah_2020,TMDC_qah_2021}.

The identification of tunable, high-Chern-number materials is of great importance for both fundamental research and practical applications.
An increased Chern number enables multiple parallel chiral edge channels, which can reduce contact resistance and enhance breakdown current~\cite{wangjing_2013,fangcheng_2014}. This property is crucial for the development of dissipationless, energy-efficient devices, as well as for multichannel topological quantum computation~\cite{Nayak_2008,alicea_2012,beenakker_2013} and low-power spintronic devices~\cite{zhangshoucheng_2012,changcuizu_2024}.

High-Chern-number QAH effects have been theoretically predicted in systems with multiple pairs of inverted subbands induced by strong exchange fields~\cite{jianghua_2012,wangjing_2013}, such as Cr-doped Bi$_2$(Se,Te)$_3$~\cite{wangjing_2013} and magnetically doped SnTe films~\cite{fangcheng_2014}. Experimentally, the Chern number in Cr-doped TI/TI multilayers can be tuned by adjusting the magnetic doping concentration or the thickness of the interior magnetic TI layers~\cite{changcuizu_2020}.

Recent theoretical predictions indicate that single-layer, stoichiometric, two-dimensional magnetic materials  are emerging as a promising platform for realizing high-temperature, high-Chern-number QAH states\cite{shengxianlei_2017,hejunjie_2017,huangchengxi_2017,wangyaping_2018,sunqilong_2018,kongxiangrui_2018,sunqilong_2019,youjiangyang_2019,sunjiaxiang_2020,qiaozhenhua_2022,
xuyong_qah_2020,sunqilong_2020,guosandong_2021,wangjing_2021,liulei_2022,guowanlin_2022,wangjing_2023,wangjing_2024}.
Monolayer M$_2$X$_2$ materials, such as Ni$_2$I$_2$, Fe$_2$Br$_2$, and Ti$_2$Te$_2$, have attracted considerable attention due to their large band gaps and high ferromagnetic transition temperatures~\cite{xuyong_qah_2020,sunqilong_2020,guosandong_2021,liulei_2022,guowanlin_2022,guo_correlation-enhanced_2022}.
An intriguing aspect of these materials is that they exhibit distinct Chern numbers despite sharing similar crystal structures.
For example, Ni$_2$I$_2$ has a Chern number of 1~\cite{liulei_2022}, whereas Fe$_2$Br$_2$ exhibits a Chern number of 2~\cite{guo_correlation-enhanced_2022}.
However, the origin of the differing Chern numbers in these materials remains largely unexplored in previous works.
This topological versatility presents an exciting opportunity for further investigation. Understanding the underlying mechanisms of this phenomenon is crucial not only for advancing our knowledge of topological phases but also for enabling the design of materials with tailored topological properties. Such control holds great promise for next-generation electronic and spintronic devices, where the unique functionalities associated with different Chern numbers could be effectively harnessed.

In this Letter, we systematically investigate monolayer M$_2$X$_2$ materials to understand the origin of their distinct Chern numbers. Our key finding is that the difference between the $C = 1$ and $C = 2$ cases lies in the location and nature of the band inversion. Specifically, the $C = 1$ system exhibits inversion at the $\Gamma$ point, whereas the $C = 2$ system features inversions along the $\Gamma$--X and $\Gamma$--Y high-symmetry lines. This contrast arises from the differing orbital compositions of the $d$-electrons near the Fermi level, which determine the symmetry properties of the involved states. This mechanism, based on band inversions at distinct $k$-points, fundamentally differs from previously proposed scenarios in which high Chern numbers result from multiple bands inversions at a single $k$-point.

\begin{figure}[tbp]
		\centering
		\includegraphics[width=0.5\textwidth]{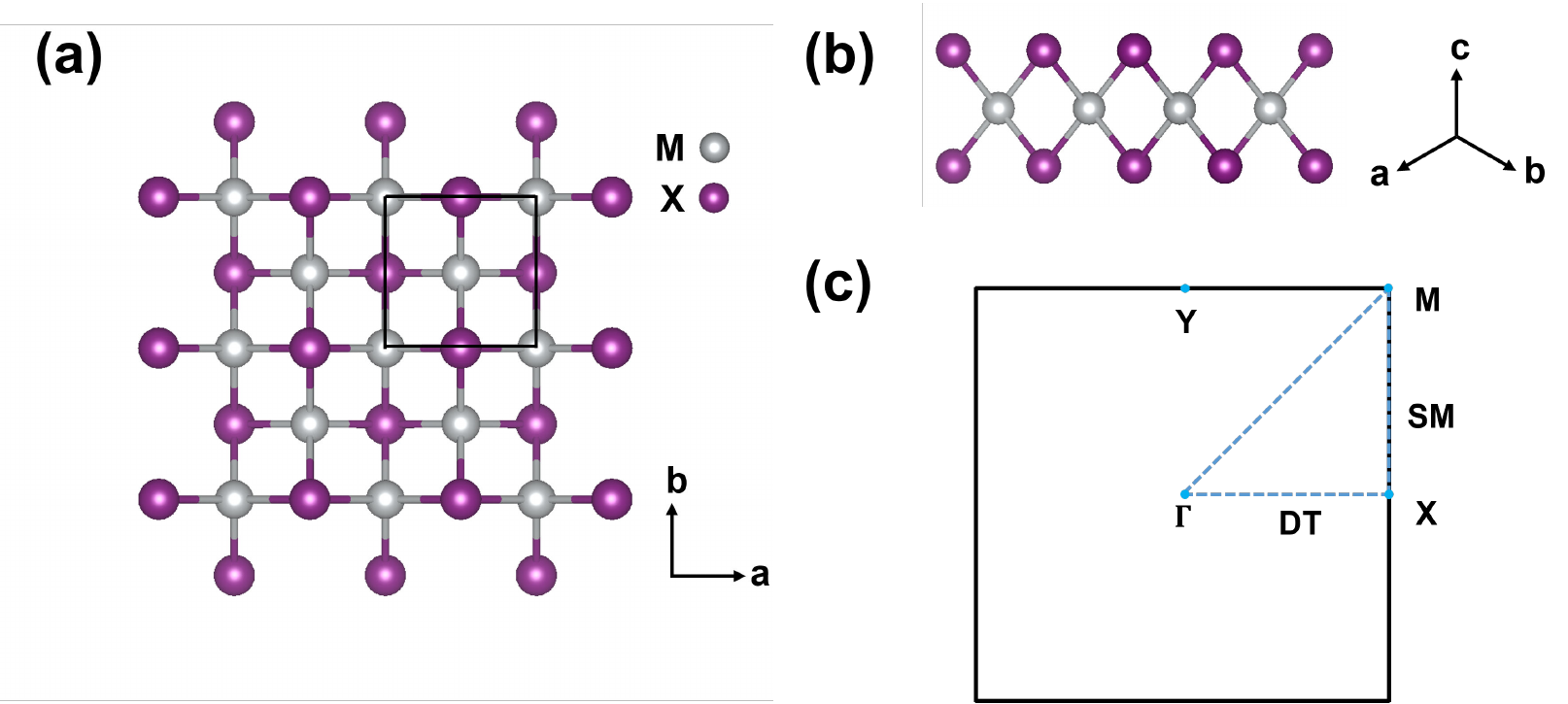}
		\caption{(a) Top view, (b) side view, and (c) Brillouin zone of monolayer M$_2$X$_2$. High-symmetry points are given in reciprocal lattice units: $\Gamma$ = (0, 0), X = (0.5, 0), Y = (0, 0.5), M = (0.5, 0.5).}
		\label{stru_MX}
	\end{figure}

The monolayer M$_2$X$_2$ has space group \textit{P}4/\textit{nmm} (No.~129).
As illustrated in Fig.~\ref{stru_MX}(a) and (b), the unit cell comprises three atomic layers, with two metal (M) atoms (denoted as M$_a$ and M$_b$ ) occupying the central layer, flanked by X atoms in the upper and lower layers.
In the top view, each M atom coordinates with X atoms from adjacent atomic layers. The X atoms occupy the 2c Wyckoff positions, while the M atoms occupy the 2b positions. Since the X atoms at 2c positions belong to different atomic layers, the tetragonal symmetry is inherently non-symmorphic:
a fourfold rotational operation alone cannot map the structure onto itself without an accompanying fractional lattice translation. Key symmetry operations also include inversion symmetry ($I$) centered at atomic positions and mirror symmetries across the $xz$ and $yz$ planes.

Given the intrinsic magnetism of the M atoms, time-reversal symmetry is inherently broken, requiring all symmetry operations to be considered within the combined real and spin space. Previous studies have shown that the M$_2$X$_2$ family exhibits ferromagnetism with high Curie temperatures and an out-of-plane easy magnetization axis~\cite{sunqilong_2020,guosandong_2021,liulei_2022,guowanlin_2022}.
When the magnetic moments of the M atoms are aligned along the $c$-axis, the system is classified under the magnetic space group
P4/{\it nm$^\prime$m$^\prime$} (No.~129.417 in the  Belov-Neronova-Smirnova (BNS) notation\cite{Gallego_db5106}).

To investigate the magnetic and electronic properties of M$_2$X$_2$, we perform first-principles calculations using the Atomic-Orbital Based Ab-initio Computation at USTC (ABACUS) software~\cite{chenmohan_2010,lipengfei_2016,linpeize_2024}. Structural relaxations and phonon calculations are carried out using the Perdew–Burke–Ernzerhof (PBE) exchange-correlation functional~\cite{pbe_1996}, while electronic band structures are computed with the Heyd–Scuseria–Ernzerhof (HSE) hybrid functional~\cite{heyd_hybrid_2003}.
Band structures with orbital projections, edge states, and anomalous Hall conductance are analyzed using the \textsc{PYATB} package~\cite{pyatb_2023}. More computational details
are presented in  Sec. A of Supplemental Material (SM)~\cite{supplementary}.

To elucidate the magnetic ordering in monolayer M$_2$X$_2$, we calculate the magnetic exchange coefficients $J$ using a magnetic force theorem implemented in TB2J package~\cite{liechtenstein_1987,tb2j_2021},  as detailed in Sec. B of  SM~\cite{supplementary}.
The results show that the exchange interactions are ferromagnetic, which aligns with the M–X–M bond angles approaching 90$^{\circ}$, as predicted by the Goodenough–Kanamori–Anderson rules~\cite{goodenough2008}.
To estimate their Curie temperatures (T$_c$), we employed Monte Carlo simulations \cite{Evans_2014} based on the Heisenberg model.
Owing to the dense packing of magnetic atoms and strong nearest-neighbor exchange interactions, the M$_2$X$_2$ monolayers exhibit notably high ferromagnetic phase transition temperatures, listed in Table \ref{table1}.

\begin{table}[tbp]
	\centering
	\caption{ First-principles calculated lattice constants ($a$); Curie temperatures (T$_c$) ; global band gaps ($E_g$) and Chern numbers ($C$) of M$_2$X$_2$ materials are listed.
Cr$_2$Bi$_2$ and Cr$_2$Sb$_2$ exhibit a local band gap but no global gap. The Chern number is obtained from the Berry curvature integral, calculated using the occupied states within a specified number of bands, rather than distinguishing by the Fermi energy.}
	\begin{ruledtabular}
		\begin{tabular*}{\columnwidth}{@{\extracolsep{\fill}} l c c c c}
				Materials  & $a$ (\AA) & T$_c$ (K)  &  $E_g$ (meV) & $C$ \\
					\hline
					Ni$_2$I$_2$   & 3.89   & 297  &  314  & 1 \\
					Ni$_2$Br$_2$  & 3.70   & 408  &  340  & 1 \\
					Cr$_2$Bi$_2$  & 4.37   & 353  &  0    &-1 \\
					Cr$_2$Sb$_2$  & 4.40   & 580  &  0    &-1 \\
					Ti$_2$Te$_2$  & 4.20   & 289  &  75   &-2 \\
					Fe$_2$I$_2$   & 3.79   & 1385 &  727  & 2 \\
					Fe$_2$Br$_2$  & 3.62   & 1531 &  293  & 2 \\
		\end{tabular*}
		\end{ruledtabular}
	\label{table1}
\end{table}


	 \begin{figure}[tbp]
	\centering
	\includegraphics[width=0.48\textwidth]{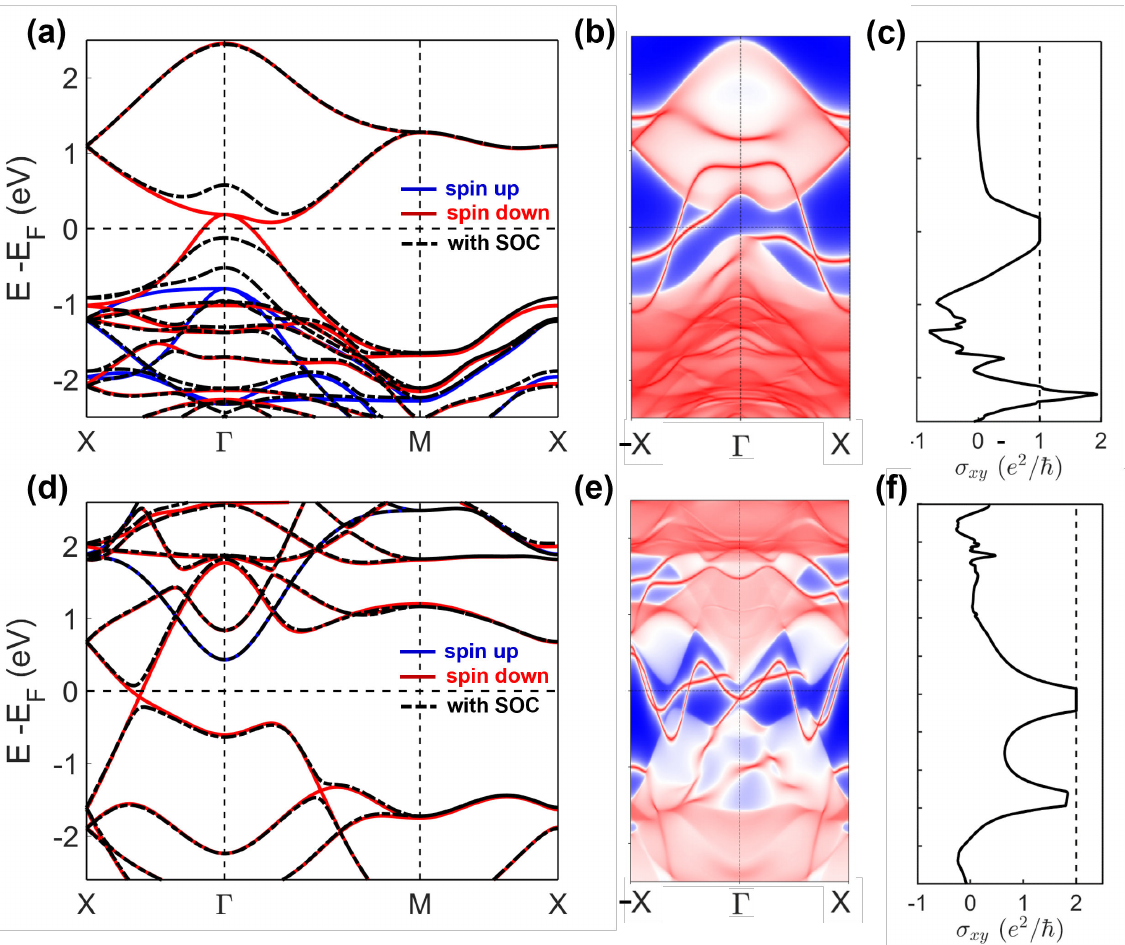}
	\caption{(a) Electronic structures of Ni$_2$I$_2$ with and without SOC; (b) topological edge states calculated along the $x$-axis; (c) anomalous Hall conductance $\sigma_{xy}$ as a function of the Fermi energy. (d)–(f): Corresponding results for Fe$_2$Br$_2$.}
	\label{band&surface}
\end{figure}

The calculated Chern numbers of the monolayer M$_2$X$_2$ family are listed in Table \ref{table1}. As shown, there are two distinct families: one with $|C|$ = 1 and the other with $|C|$ = 2.
To investigate the origin of distinct Chern numbers in M$_2$X$_2$, we select Ni$_2$I$_2$\cite{liulei_2022} and Fe$_2$Br$_2$\cite{guo_correlation-enhanced_2022} as representative materials, possessing Chern numbers of 1 and 2, respectively.
Figure~\ref{band&surface}(a) and (d) present the band structures of Ni$_2$I$_2$ and Fe$_2$Br$_2$ calculated with/without
spin-orbit coupling (SOC).
When the SOC is turned off, in Ni$_2$I$_2$, the conduction band (CB) and valence band (VB) touch at the $\Gamma$ point, whereas in Fe$_2$Br$_2$, a band crossing emerges along the high-symmetry directions from the $\Gamma$ point to the X and Y points due to the fourfold rotational symmetry.
However, when SOC is included, both materials open energy gaps at the band crossing points near the Fermi level.
Especially, Ni$_2$I$_2$ and Fe$_2$Br$_2$ exhibit substantial band gaps of 314 meV and 293 meV, respectively, as calculated with the HSE  functional~\cite{heyd_hybrid_2003}.
These values are significantly larger than those of Bi$_2$Se$_3$\cite{zhang_topological_2009} and MnBi$_2$Te$_4$\cite{wangjing_MBT_2019}.

 We further examine the edge states of the two materials on the ribbon and calculate the variation of Hall conductance with respect to the Fermi energy. The edge-state calculations~\cite{sancho_quick_1984} in Fig.~\ref{band&surface}(b) reveal three bands crossing the Fermi level, two with positive slopes and one with a negative slope, indicating that Ni$_2$I$_2$ is a Chern insulator with a $C$=1. This is further supported by the quantized conductance plateau of $e^2/h$ at the Fermi level, as shown in Fig.~\ref{band&surface}(c).
Similarly, edge-state calculations in Fig.~\ref{band&surface}(e) and Hall conductance calculations in Fig.~\ref{band&surface}(f) demonstrate that Fe$_2$Br$_2$ is a Chern insulator with $C = 2$.
These results are consistent with previous reports\cite{liulei_2022,guo_correlation-enhanced_2022}.
Detailed band structures and surface state calculations for the additional materials are provided in Sec. C of  SM~\cite{supplementary}.

\begin{figure}[t]
	\centering
	\includegraphics[width=0.5\textwidth]{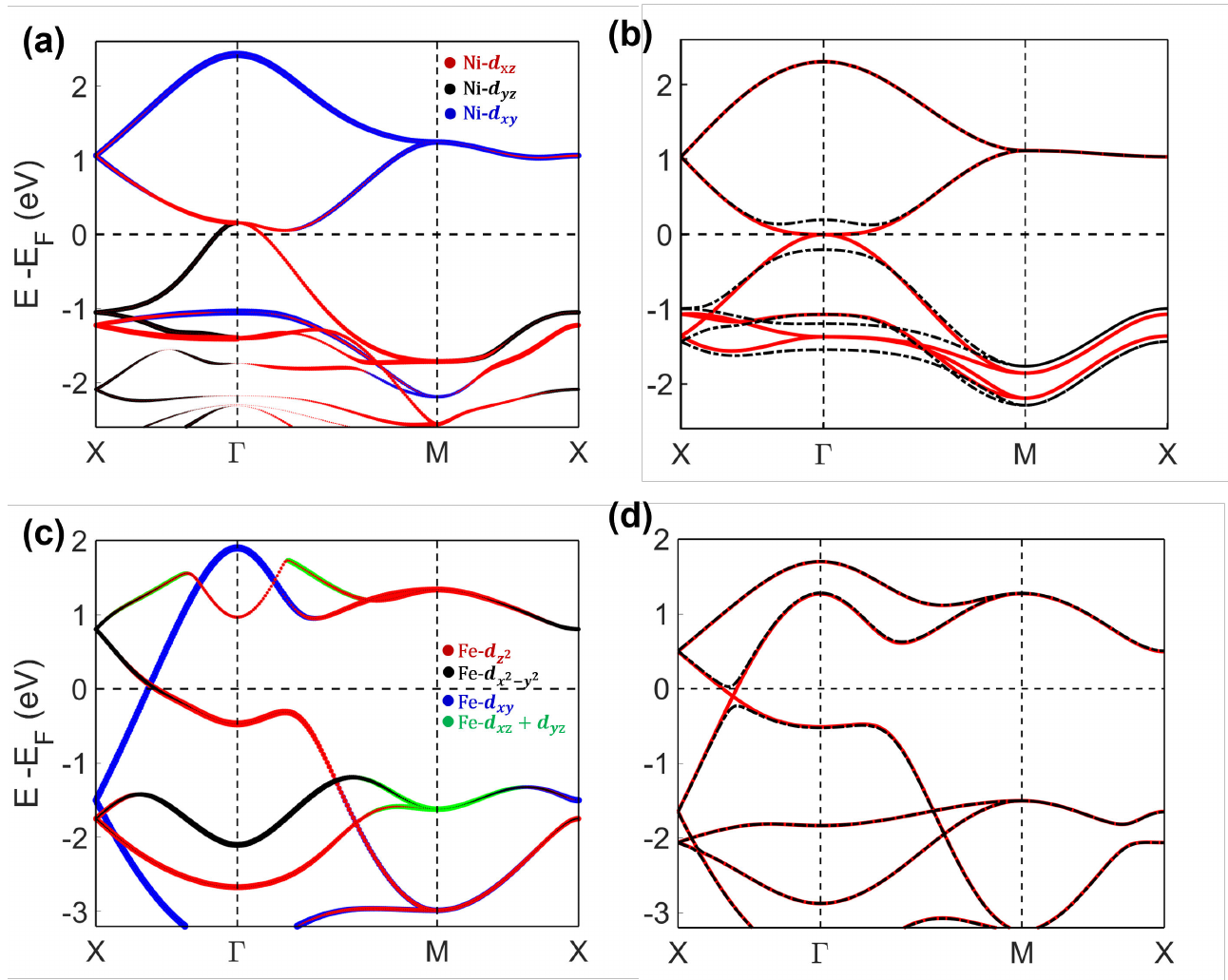}
	\caption{(a) The $d$-orbital projected band structures of monolayer Ni$_2$I$_2$ without SOC. The sizes of the circles indicate the weight of different $d$ orbitals. (b) The tight-binding model bands without (red solid lines) and with (black dashed lines) SOC. (d) and (e) show the corresponding results for Fe$_2$Br$_2$.}
	\label{fatband&tb}	
\end{figure}

Figure~\ref{fatband&tb}(a) and (c) show the orbital contributions to the band structures without SOC for Ni$_2$I$_2$ and Fe$_2$Br$_2$, respectively.
In both materials, the states close to the Fermi level are predominantly composed of spin-down $d$-orbital electrons.
According to Hund's rule, Ni atoms in Ni$_2$I$_2$ possess five spin-up and four spin-down electrons in their $d$-orbitals, resulting in a magnetic moment of 1~$\mu_B$,  which is consistent with first-principles calculations.
In contrast, Fe atoms in Fe$_2$Br$_2$ have five spin-up electrons and two spin-down electrons, corresponding to a magnetic moment of 3~$\mu{B}$.
As we shall see  the distinct electron occupations lead to significant differences in their properties.

To provide a clearer physical interpretation of the topological properties, we construct tight-binding (TB) models to study the electronic structure of the two materials. In an octahedral crystal field, the $d$-orbitals split into a two-fold degenerate $\text{E}_\text{g}$ states ($d_{x^2 - y^2}$, $d_{z^2}$) and a three-fold degenerate $\text{T}_{2\text{g}}$ states ($d_{xy}$, $d_{xz}$, $d_{yz}$). The $\text{E}_\text{g}$ states are lower in energy because their orbitals point directly toward the negatively charged ligands. When the symmetry is lowered to tetrahedral, the $\text{E}_\text{g}$ degeneracy splits into the nondegenerate $\text{A}_1$ and $\text{B}_1$ representations, while the $\text{T}_{2\text{g}}$ level splits into a nondegenerate $\text{B}_2$ and a two-fold degenerate $\text{E}$ representation~\cite{supplementary}.

As shown in Fig.~\ref{fatband&tb}(a), the bands near the crossing points in Ni$_2$I$_2$ predominantly involve the $d_{xz}^{\downarrow}$, $d_{yz}^{\downarrow}$, and $d_{xy}^{\downarrow}$ orbitals, originating from the $\text{T}_{2\text{g}}$ manifold. The corresponding TB model includes these three $d$-orbitals, along with the I $p_{x}^{\downarrow}$ and $p_{y}^{\downarrow}$ orbitals, resulting in a ten-orbital basis per unit cell, which comprises two Ni atoms and two I atoms.
In contrast, due to the difference in electron occupation, the relevant bands in Fe$_2$Br$_2$ mainly consist of the $d_{z^2}^{\downarrow}$, $d_{xy}^{\downarrow}$, and $d_{x^2 - y^2}^{\downarrow}$ orbitals, which include both $\text{E}_\text{g}$ states and orbitals from $\text{T}_{2\text{g}}$ states.
In this material, the TB model involves exclusively the Fe $d$-orbitals, and the Br $p$ orbitals can be neglected because their contribution is minimal near the Fermi level.
Both TB models incorporate nearest-neighbor and next-nearest-neighbor hopping processes and are analyzed both with and without SOC.
More details of the TB Hamiltonians are provided in the Section D of SM~\cite{supplementary}.

Figure~\ref{fatband&tb}(b),(d) show band structures calculated from TB models for Ni$_2$I$_2$ and Fe$_2$Br$_2$ respectively,  which closely reproduce the band dispersions and irreducible representations obtained from first-principles calculations in Fig.~\ref{band&surface}.
More importantly, the TB models reproduce the Chern numbers of the two materials, which agree with the first-principles calculations (see SM~\cite{supplementary}).

The essence of topological classification lies in the fact that topological materials cannot be expressed as a sum of elementary band representations (EBRs)~\cite{po_symmetry-based_2017,bradlyn_topological_2017,zhang_catalogue_2019,tang_comprehensive_2019,elcoro_magnetic_2021}.
Here, we offer a clear and physically intuitive explanation for the emergence of different Chern numbers in magnetic M$_2$X$_2$ materials. From a symmetry standpoint, different $d$ orbitals correspond to distinct EBRs at high-symmetry $\mathbf{k}$ points in the Brillouin zone, as summarized in Table~\ref{band representations}.

At the X/Y points, symmetry constraints cause the effective interactions between M$_a$ and M$_b$, which occupy the same Wyckoff site, to vanish. Symmetry operations interchange the M$_a$ and M$_b$ sites, resulting in at least doubly degenerate bands at these points for different $d$ orbitals.
 In contrast, at the $\Gamma$ point, strong interactions between the M$_a$ and M$_b$ atoms lift this degeneracy, causing the sublattice-symmetric orbitals, previously degenerate at the X/Y points, to split into bonding and antibonding states. When the interaction between different orbitals of the M$_a$ and M$_b$ atoms exceeds the on-site energy difference at the X/Y points, a band inversion may occur along the $\Gamma \rightarrow$ X/Y high-symmetry lines.
Depending on the orbitals' arrangement near the Fermi level, two distinct scenarios can arise.

\begin{table}[tbp]
\caption{$d$-orbital band representations at X, $\Gamma$, and M for space group $P4/nm'm'$ (without time-reversal symmetry). The numbers in parentheses indicate the degeneracy of each representation.}
	\begin{ruledtabular}
		\begin{tabular}{c|c|c|c}
			& X  & $\Gamma$  & M \\ \hline
			$d_{z^2}@2b$        & X$_1(2)$                  & G$_{1}^{+}(1)$ $\oplus$  G$_{4}^{-}(1)$    & M$_2(2)$ \\
			$d_{x^2-y^2}@2b$    & X$_1(2)$                  & G$_{2}^{+}(1)$ $\oplus$  G$_{3}^{-}(1)$    &M$_1(2)$  \\
			$d_{xy}@2b$         & X$_2(2)$                  & G$_{3}^{+}(1)$ $\oplus$  G$_{2}^{-}(1)$    &M$_1(2)$  \\
			$d_{xz,yz}@2b$      &  X$_1(2)\oplus$X$_2(2)$   & G$_{5}^{+}(2)$ $\oplus$  G$_{5}^{-}(2)$    & M$_3(2)$ $\oplus$ M$_4(2)$ \\
		\end{tabular}
	\end{ruledtabular}
	\label{band representations}
\end{table}

\begin{figure}[tb]
	\centering
	\includegraphics[width=0.4\textwidth]{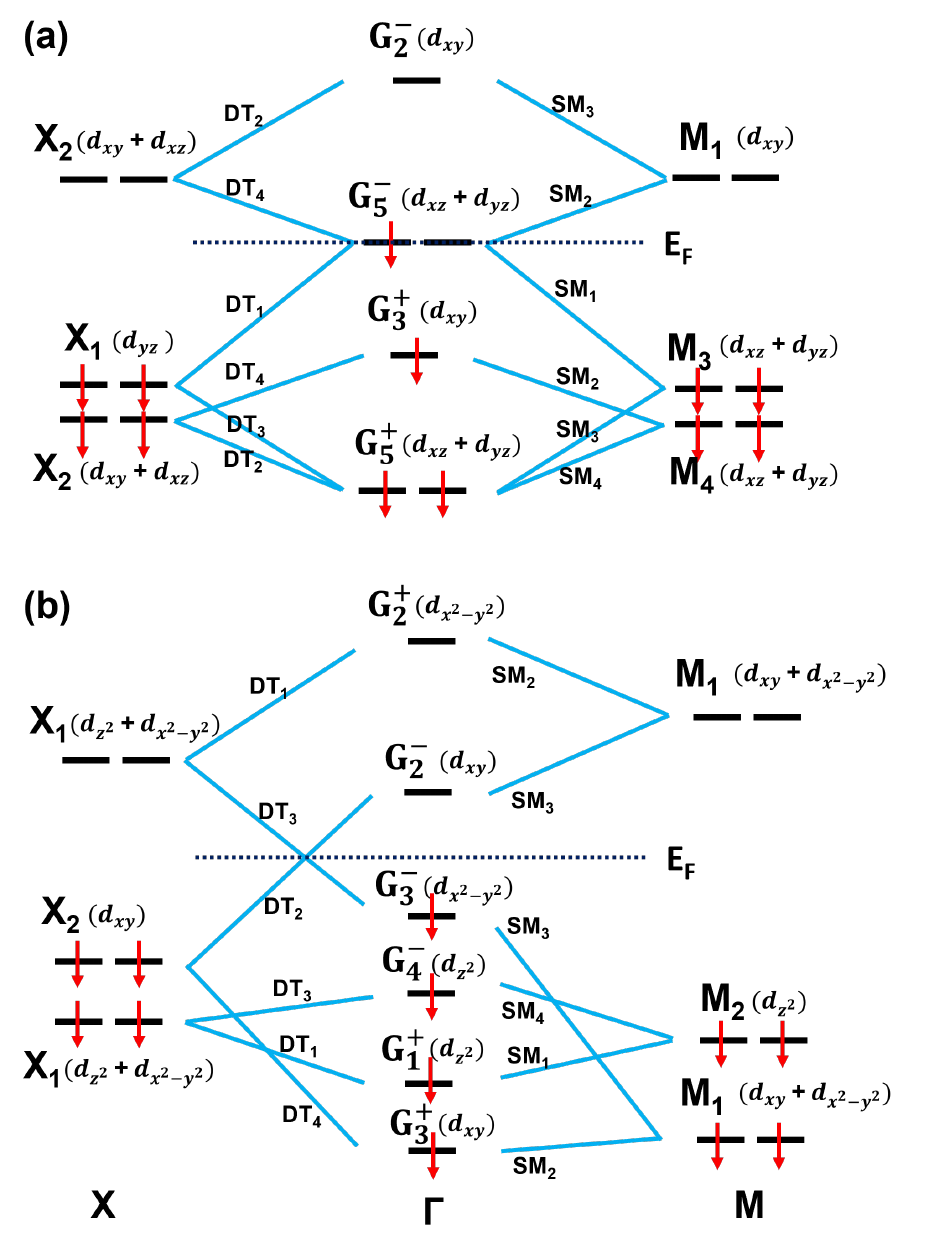}
	\caption{The band connectivity near the Fermi level without SOC for monolayer (a) Ni$_2$I$_2$ and (b) Fe$_2$Br$_2$ is shown along the X $\rightarrow$ $\Gamma$ $\rightarrow$ M path. At each high-symmetry point and along high-symmetry lines, the corresponding irreducible representation is indicated, with the orbital character at the high-symmetry points specified in parentheses.}
\label{band_connection}	
\end{figure}

The first scenario is depicted in Fig.~\ref{band_connection}(a) for Ni$_2$I$_2$. At the X point, the unoccupied X$_2$ band is dominated by $d_{xy}$ orbitals with small potion of $d_{xz}$ orbtials, the X$_1$ state is primarily composed of $d_{yz}$ orbitals, and the lower-energy X$_2$ band is mainly derived from $d_{xz}$ orbitals with small potion of $d_{xy}$ orbtials . All bands are sublattice-degenerate.
At the $\Gamma$ point, strong M$_a$–M$_b$ interactions split the $d_{xy}$ orbital into bonding ($\text{G}_3^{+}$) and antibonding ($\text{G}_2^{-}$) states, while the $d_{xz}$/$d_{yz}$ orbitals form doubly degenerate bonding ($\text{G}_5^{+}$) and antibonding ($\text{G}_5^{-}$) states. The $\text{G}_3^{+}$ band lies below the $\text{G}_5^{-}$ band, indicating a band inversion.
The twofold-degenerate $\text{G}_5^{-}$ state at the $\Gamma$ point serves as a band crossing in the absence of SOC, indicating that the occupied states cannot be expressed as a sum of EBRs derived from the $d_{xz}/d_{yz}$ orbitals. When SOC is included, a band gap opens at $\Gamma$ between the $\text{G}_5^{-}$ states.

 The topological properties of band structures can be diagnosed from the wave functions at high-symmetry $\mathbf{k}$-points in the Brillouin zone~\cite{fuliang_2007,fangchen_2012}.
We compute the symmetry-protected topological indices for Ni$_2$I$_2$ using the tight-binding model, based on the symmetry indicator theory~\cite{po_symmetry-based_2017,bradlyn_topological_2017,zhang_catalogue_2019,tang_comprehensive_2019,elcoro_magnetic_2021}.
The magnetic space group of M$_2$X$_2$ is generated by the symmetry operations $C_{4z}$, inversion $I$, and mirror reflection $m_{001}$, along with their products.
For Ni$_2$I$_2$, we obtain a symmetry indicator $\text{Z}_{4\text{R}} = 1$. We further evaluate additional symmetry indicators associated with $C_{2z}$, inversion ($I$), and $S_{4z} = C_{4z} \cdot I$, finding that $\text{Z}_{2\text{R}}$, $\text{Z}_{2\text{I},3}$, and $\text{Z}_{4\text{S}}$ are all equal to 1. These topological indicators, consistent with the first-principles result $C = 1$, provide valuable insights into the symmetry-enforced origin of the topological phase.
Further details are provided in the SM~\cite{supplementary}.

The second scenario is depicted in Fig.~\ref{band_connection}(b) for Fe$_2$Br$_2$. At the X point, the lower-energy X$_1$ band is primarily derived from the $d_{z^2}$ orbital, the higher-energy X$_1$ band mainly from the $d_{x^2 - y^2}$ orbital, and the X$_2$ band originates from the $d_{xy}$ orbital.
At the $\Gamma$ point, these doubly degenerate orbitals split into bonding and antibonding states.
In particular, strong M$_a$–M$_b$ interactions induce a band inversion: the antibonding state derived from the $d_{xy}$ orbital ($\text{G}_2^{-}$) is pushed above the antibonding state associated with the $d_{x^2 - y^2}$ orbital ($\text{G}_3^{-}$), resulting in an unoccupied $\text{G}_2^{-}$ and an occupied $\text{G}_3^{-}$ state.
Therefore, the occupied states cannot be represented as EBRs of $d_{z^2}/d_{xy}$ orbitals.
According to compatibility relations, the DT$_2$ representation, connecting X$_2$ with $\text{G}_2^{-}$, and the DT$_3$ representation, connecting X$_1$ with $\text{G}_3^{-}$, along the high-symmetry path, must exhibit an accidental band crossing. When SOC is included, a band gap opens at the crossing point along the high-symmetry line.
Due to $C_{4z}$ symmetry, a similar band inversion also occurs along the $\Gamma$–$\bar{\rm X}$ and $\Gamma$–Y$ (\bar{\rm Y}$)  paths, resulting in a total of four crossing points. Each crossing point contributes 1/2 to the total Chern number, as can be easily seen from a $\mathbf{k}$$\cdot$$\mathbf{p}$ calculation, leading to a Chern number of 2 for monolayer Fe$_2$Br$_2$ (see Sec. F of the SM~\cite{supplementary}).
We also calculate the topological symmetry indicators and obtain $\text{Z}_{4\text{S}} = \text{Z}_{4\text{R}} = 2$ and $\text{Z}_{2\text{R}} = \text{Z}_{2\text{I},3} = 0$.


 Based on the above analysis, the primary reason for the different Chern numbers in M$_2$X$_2$ materials lies in the distinct symmetries and fillings of the $d$ orbitals near the Fermi level.
If the dominant orbitals are $d_{xz}$ and $d_{yz}$, a two-dimensional (doubly degenerate) representation emerges at the $\Gamma$ point (in the absence of SOC), leading to a band inversion at $\Gamma$ and resulting in a Chern number of $C = 1$.
In contrast, if other orbitals dominate, the band crossing points occur along the $\Gamma$–X line and, by symmetry, also along the $\Gamma$–Y line.
This yields four symmetry-related crossing points protected by $C_{4z}$ symmetry, giving rise to $C = 2$.

For monolayer Ni$_2$Br$_2$, Cr$_2$Bi$_2$, and Cr$_2$Sb$_2$, the band characteristics near the Fermi level are similar to those of Ni$_2$I$_2$, with the $d_{xz}$ and $d_{yz}$ orbitals dominating.
In the Cr-based systems, however, the $d$ orbitals are less than half-filled, and the states near the Fermi level are predominantly from spin-up electrons, leading to a Chern number of $C = -1$.
Detailed analyses of the orbital composition and topological edge states are presented in Sec. C of
SM~\cite{supplementary}.

Recently, monolayer PdTaX$_2$ (X = Se, Te) structures~\cite{he_excellent_2024}  have been predicted to possess a Chern number of 1.
In these materials, the Ta $d_{xz}$ and $d_{yz}$ orbitals dominate near the Fermi level, and a band inversion occurs at the M point.
By constructing a $\sqrt{2} \times \sqrt{2}$ supercell, the arrangement of Ta atoms becomes equivalent to that of the M atoms in the M$_2$X$_2$ structure~\cite{supplementary}.
This supercell transformation folds the M point onto the $\Gamma$ point, producing a band structure similar to that of Ni$_2$I$_2$.
Accordingly, the origin of the Chern number $C = 1$ in monolayer PdTaX$_2$ can be understood within the theoretical framework established above(see Sec. G of the SM~\cite{supplementary}).

Within the M$_2$X$_2$ family, Fe$_2$I$_2$ shares the same band structure as Fe$_2$Br$_2$, with near-Fermi states dominated by spin-down $d$ orbitals other than $d_{xz}$ and $d_{yz}$, yielding $C = 2$.
A slightly different case is Ti$_2$Te$_2$~\cite{guowanlin_2022}, where the overall orbital symmetry near the Fermi level remains similar, but the $d$ shell is less than half-filled.
In this case, the near-Fermi states are mainly spin-up, reversing the Berry-curvature sign and giving $C = -2$.

This $C = 2$ mechanism also applies beyond the M$_2$X$_2$ framework.
For example, in LiFeSe~\cite{xuyong_qah_2020}, interstitial Li donates one electron, producing an orbital composition near the Fermi level that closely matches Fe$_2$Br$_2$ and thus also yields $C = 2$.
KTiSb~\cite{wangjing_2023} and MgFeP~\cite{yao_orbital-selectivity-induced_2024} follow the same principle.
Similarly, Janus monolayers~\cite{guo_intrinsic_2021} derived from M$_2$X$_2$ preserve the near-Fermi orbital character when the two chalcogen/halogen layers are chosen from the same group (e.g., Fe$_2$IX (X = Cl, Br)\cite{guo_intrinsic_2021}), thereby retaining the same Chern number.


To summarize, we have systematically analyzed the topological properties of monolayer M$_2$X$_2$ materials, clarifying the origin of their different Chern numbers. We find that differences in the 3$d$ orbital composition and symmetry representations near the Fermi level give rise to two primary band-inversion mechanisms, which in turn generate distinct Chern numbers and topological phases. In particular, one mechanism, governed by crystal symmetry (e.g., $C_{4z}$), allows band inversions to occur concurrently at multiple symmetry-related $\mathbf{k}$ points across the Brillouin zone. This leads to Berry-curvature contributions that accumulate to yield higher Chern numbers, a scenario distinct from earlier mechanisms that relied on multiple band inversions at a single $\mathbf{k}$ point. This insight provides clear criteria for screening and engineering monolayer materials with tailored topological properties, particularly high-Chern-number systems. Our results establish a robust theoretical framework and offer practical guidance for realizing quantum anomalous Hall insulators with large band gaps and elevated ferromagnetic transition temperatures, suitable for advanced spintronics and topological quantum computing.

This work was supported by the National Natural Science Foundation of China (Grant Number 12134012),
the Strategic Priority Research Program of Chinese Academy of Sciences (Grant Number XDB0500201),
and the Innovation Program for Quantum Science and Technology Grant Number 2021ZD0301200.
The numerical calculations were performed on the USTC HPC facilities.

%

\end{document}